# STREAMING MOTION IN LEO I

By Bodhisattva Sen,[1] Moulinath Banerjee,[2] Michael Woodroofe,[3] Mario Mateo and Matthew Walker

*University of Michigan, University of Michigan, University of Michigan, University of Michigan, and University of Cambridge*

Whether a dwarf spheroidal galaxy is in equilibrium or being tidally disrupted by the Milky Way is an important question for the study of its dark matter content and distribution. This question is investigated using 328 recent observations from the dwarf spheroidal Leo I. For Leo I, tidal disruption is detected, at least for stars sufficiently far from the center, but the effect appears to be quite modest. Statistical tools include isotonic and split point estimators, asymptotic theory, and resampling methods.

**1. Introduction.** The dwarf spheroidal galaxies near the Milky Way are among the least luminous galaxies in the night sky. While they have stellar populations similar to those of globular clusters, approximately $10^6$–$10^7$ stars, they are considerably larger systems, typically hundreds, even thousands of parsecs in size compared to radii of tens of parsecs characteristic of clusters. They are excellent candidates for the study of dark matter because they are nearby and generally have extremely low stellar densities. Moreover, due to their proximity to the Milky Way, many of these dwarf galaxies are also potentially strongly affected by disruptive tidal effects. The mere existence of dwarf spheroidal galaxies suggests these systems contain much dark matter since it is unclear how they could have avoided the effects of tidal disruption without the added gravitational force from a considerable reservoir of unseen matter [see Muñoz, Majewski and Johnston (2008) and Mateo, Olszewski and Walker (2008)].

Detailed kinematic studies [Wang et al. (2005), Walker et al. (2006) and Wang et al. (2008a)] confirm the widespread belief that the dwarf spheroidal

Received June 2008; revised August 2008.
[1]Supported in part by NSF Grants AST-05-07453 and DMS-07-05288.
[2]Supported in part by NSF Grant DMS-07-05288.
[3]Supported in part by NSF Grant AST-05-07453.
*Key words and phrases.* Asymptotic distributions, bootstrap, change point models, dark matter, dwarf spheroidal galaxies, isotonic estimation, permutation tests, split points.

galaxies are dominated by dark matter, in many cases finding that the dark matter densities exceed that of visible matter by a few orders of magnitude. These latest studies differed from their predecessors, for example, Mateo et al. (1993), Mateo (1998), Mateo et al. (1998) (and references therein), and Kleyna et al. (2002), statistically by using a nonparametric analysis to estimate the distribution of dark matter. While the nonparametric analysis did not require a specific form for this distribution, it did assume that the galaxies are in equilibrium and isotropic. The purpose of the present paper is to explore the gravitational effects of the Milky Way on the dwarf spheroidals and, implicitly, to probe the underlying assumptions used in Wang et al. (2005), Walker et al. (2006) and Wang et al. (2008a).

To this end, our approach is to address these issues using recent data [Mateo, Olszewski and Walker (2008)] for the dwarf spheroidal galaxy Leo I from which we obtain kinematic observations of 328 stars. Among the Milky Way's dSph satellites, Leo I is perhaps the most distant, at $255 \pm 3$ kpc, and is receding from the Milky Way at a relatively large velocity of $179.5 \pm 0.5$ km/s. The combination of the large distance and high velocity led Byrd et al. (1994) to suggest that Leo I may not be bound to the Milky Way. Bound or unbound, the large outward velocity means that Leo I passed much closer to the Galactic Center in the past. In the preferred model of Byrd et al. (1994), Leo I passed within 70 kpc of the Galactic Center, a distance similar to the closest present day dwarf spheroidal galaxies. Recent papers [Sohn et al. (2007), Mateo, Olszewski and Walker (2008)] suggest more specific models in which Leo I passed within 10–20 kpc from the center of the Milky Way some 1–2 Gyr ago.

So, the question becomes as follows: What effect, if any, did this close encounter have on Leo I? In some cases, a close encounter with the Galactic Center can change the shape of a dwarf spheroidal by producing tidal arms [Oh, Lin and Aarseth (1995), Piatek and Pryor (1995)]. Prominent tidal arms are not observed in Leo I, but this may reflect our unfavorable viewing angle rather than the actual lack of such features [Mateo, Olszewski and Walker (2008)]. A more subtle but related effect is streaming motion. The practical observational signal of this process arises from the fact that both leading and trailing stars move away from the center of the main body of the perturbed systems in the reference frame of that galaxy. Since stars near the center are more tightly bound to the galaxy than stars near the edge, the magnitude of streaming is expected to increase with distance from the center and be aligned with the apparent major axis of the galaxy. Another question of interest here is whether there is a threshold below which the magnitude of streaming motion is negligible and beyond which it is appreciable, as suggested by Mateo, Olszewski and Walker (2008). A threshold model is motivated by the nature of the tidal forces a dwarf galaxy experiences as it follows a highly elliptical orbit around the Milky Way. A more detailed account of streaming motion may be found in Section 4.3 of Mateo, Olszewski and Walker (2008).

There are several interesting statistical questions here. Is streaming motion evident in Leo I? If so, how can it be described and estimated? To what extent can it be described by a threshold model, in which streaming motion is only present for stars at a sufficient distance from the center? We answer these questions within the context of a model, called the cosine model below, that incorporates the qualitative features of streaming motion described above (increasing with distance from the center and largest along the major axis). The answers may be summarized: The magnitude of streaming motion appears to be modest, at most 6.19 km/s, but is (nearly) significant at the 5% level. The streaming motion does appear to be consistent with a threshold model, but it is difficult to constrain the threshold. This may reflect the inherent "fuzziness" of such a threshold radius, plus the fact that, due to projection effects, stars associated with streaming motions can be superposed on the sky with regions of stars that do not show any streaming.

The data are described in Section 2. In Section 3 we review the bisector test used in Mateo, Olszewski and Walker (2008). The cosine model is introduced in Section 4 and used to estimate the magnitude of streaming motion and motivate a test for significance. Threshold models are considered in Section 5. Section 6 contains remarks, outlining possible extensions. The Appendix states some of the asymptotic results used in the paper.

## 2. The data.

*The data.* The data used here consist of position and velocity measurements for candidate member stars from Leo I. These were derived from observations using the multi-fiber Hectochelle spectrograph on the MMT telescope at Las Campanas Observatory during March and April of 2005, 2006, and 2007. The raw spectra were converted to velocity measurements using **fxcor** in IRAF (the Image Reconstruction and Analysis Facility), which returns a velocity measurement and an estimate of the standard deviation of measurement error for each star. A detailed description of the observation and reduction processes is included in Mateo, Olszewski and Walker (2008). For each star the four variables of primary interest here were line of sight velocity, position projected on the plane orthogonal to the line of sight, and the standard deviation of measurement error for the velocity. Velocities $Y$ and the standard deviations $\Sigma$ are expressed in km/s. Position is expressed in polar coordinates $(R, \Theta)$, with $R$ measured in arc seconds and $\Theta$ in degrees, so defined that $\Theta = 0$ or 180 along the major axis. For Leo I, 400 arc seconds are roughly 500 parsecs.

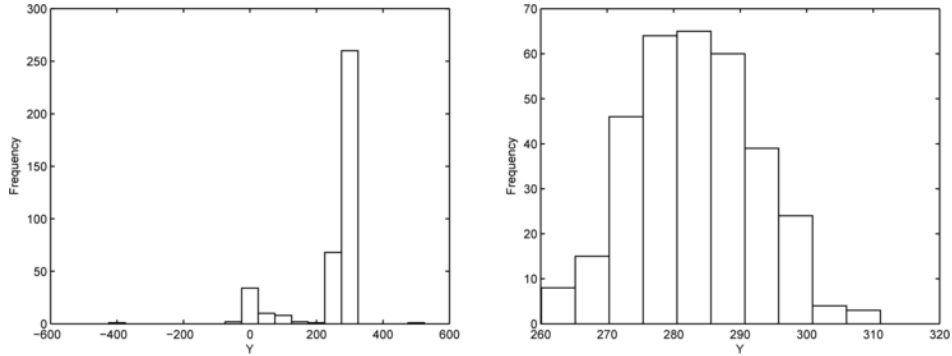

Fig. 1. *Histogram of velocities for Leo I before and after trimming.*

*Trimming.* A complicating feature of the data is that not all stars in the sample are really members of the galaxy. Some are foreground stars, located along our line of sight toward Leo I. Fortunately, due to Leo I's large systemic velocity, the radial velocities of nonmembers are quite distinct from those of the galaxy itself. Velocities of the galaxy members are fairly tightly clustered around a well-defined and, in this case, very large positive velocity, while velocities of nonmembers have a much broader distribution centered much closer to zero heliocentric velocity; see Figure 1. To eliminate the foreground stars, we computed (an estimate of) the probability that each star is a galaxy member, using the method of Sen, Walker and Woodroofe (2008). For the Leo I, these probabilities were either at least 0.99 or at most 0.01. We eliminated the stars with low probabilities and kept 328 others. Some descriptive statistics of the trimmed sample are presented in Table 1. Observe that the trimmed sample consists of stars whose velocities are within three standard deviations of their mean. The data from Leo I is exceptional in that *all* of the inclusion probabilities are either very high or very low. Data from the dwarf spheroidals Sculptor and Sextans provide examples with many intermediate values. In such cases, it is sometimes possible to include other variables, like metallicity (magnesium and calcium indices), to improve discrimination [see Walker et al. (2008)].

*Selection.* The data are regarded as a random sample from Leo I, but not a simple random sample, since some regions were sampled more extensively than others. Thus, the joint density of $(R, \Theta)$ is of the form

$$(1) \qquad f(r, \theta) \propto u(r, \theta) g(r, \theta),$$

where $g$ is the density of $R$ and $\Theta$ within the population and $u$ is the selection function. Figure 2 presents a scatter plot of $(R, \Theta)$ for the trimmed sample from which some effects of selection may be seen: Within the population of

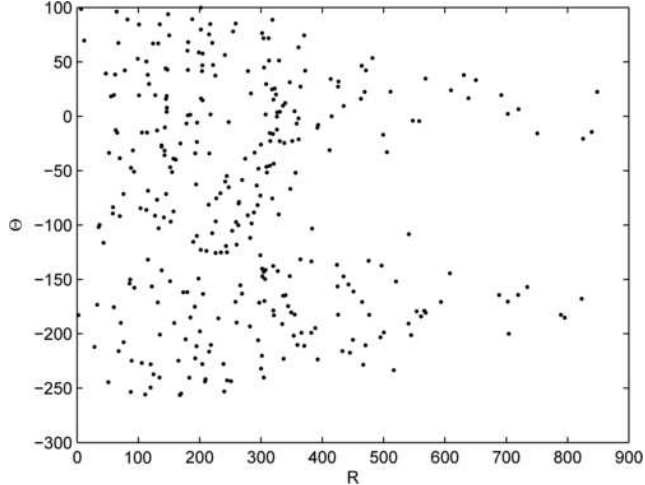

Fig. 2. *Scatterplot of $(R, \Theta)$ for the Leo* I *data.*

Leo I stars, it is not unreasonable to assume that $R$ and $\Theta$ are independent and that $\Theta$ has a uniform distribution [Mateo, Olszewski and Walker (2008)]. The data in Figure 2 are clearly not consistent with this assumption: Stars beyond 400 arc seconds divide into two groups, one centered roughly around $\theta = 0$ and the other roughly around $\theta = -180$. That is, for stars beyond 400 arc seconds, the sample divides into two branches centered around the major axis on the two sides of the galaxy. This assumption was not intentional since candidate members were selected as uniformly as feasible in $\Theta$ over the full range of $R$ shown in Figure 2.

If we do suppose that $R$ and $\Theta$ are independent and that $\Theta$ has a uniform distribution, within the population, then it is possible to estimate the selection function $u$. The marginal density of $R$ within the population of Leo I stars can be estimated with some precision from the large sample of positions reported in Irwin and Hatzidimitriou (1995). Thus, the joint density $g$ of $R$ and $\Theta$ can be estimated with some precision. It is also possible to estimate $f$ from our selected sample, using a kernel estimate, for example, and then $u$ can be recovered from (1). Calculations of this nature are reported in Wang et al. (2005). We do not pursue this here because most of our analysis is conditional on position and, so, unaffected by the selection.

TABLE 1
*Descriptive statistics for Leo* I

|  | R | $\Theta$ | $\cos(\Theta)$ | Y | $\Sigma$ |
|---|---|---|---|---|---|
| min | 2.3000 | $-256.3000$ | $-1.0000$ | 260.1000 | 1.6000 |
| max | 848.5000 | 99.8000 | 1.0000 | 311.1000 | 7.6000 |
| med | 259.8000 | $-74.2000$ | 0.0932 | 282.6500 | 2.0000 |
| mean | 283.3213 | $-77.6591$ | 0.0932 | 283.0927 | 2.1302 |
| stdev | 171.9573 | 100.4565 | 0.7619 | 9.4144 | 0.6470 |

*A Model.* To describe the effects of streaming motion, let $V$ denote the line of sight velocity of a star and suppose that, within a galaxy, $R$ and $\Theta$ have a joint density $g$ and $V = \nu(R, \Theta) + \varepsilon$, where $\varepsilon$ is a random fluctuation with mean 0 and variance $\sigma^2$, and $\varepsilon$ is independent of $(R, \Theta)$. Thus, $\nu(r, \theta)$ is the expected velocity, given $R = r$ and $\Theta = \theta$. Velocity is measured with some error. We observe $(Y, \Sigma)$, where $Y = V + \delta$ and the conditional distribution of $\delta$ given $(R, \Theta, \varepsilon, \Sigma)$ is normal with mean 0 and standard deviation $\Sigma$. Thus, for the selected sample, $(R_i, \Theta_i, Y_i, \Sigma_i)$, $i = 1, \ldots, n = 328$, are independent and identically distributed random vectors for which $(R_i, \Theta_i)$ have density $f$,

$$(2) \qquad Y_i = \nu(R_i, \Theta_i) + \varepsilon_i + \delta_i,$$

where $(R_i, \Theta_i), \varepsilon_i$, and $\Sigma_i$ are independent, and the conditional distribution of $\delta_i$ given $(R_i, \Theta_i, \varepsilon_i, \Sigma_i)$ is normal with mean 0 and variance $\Sigma_i^2$. The standard deviation $\sigma$ is called the velocity dispersion and is of interest because of its relationship to the mass distribution. Every known (to us) estimator of galaxy mass uses an estimate of the velocity dispersion at some level. These range from simple expressions, like $\hat{\sigma}^2/r$, for a disk galaxy, to more complex relations, like Jeans' Equation, for pressure driven systems, like Leo I; see Binney and Tremaine (1987).

For the remainder of the paper, let $r_1 \leq r_2 \leq \cdots \leq r_n$ denote the ordered values of $R_1, \ldots, R_n$, and let $\theta_1, \ldots, \theta_n$, $\sigma_1, \ldots, \sigma_n$, and $y_1, \ldots, y_n$, the concomitant order statistics of $\Theta_i$, $\Sigma_i$, and $Y_i$. To avoid selection effects, we condition on the position variables $r_1, \ldots, r_n, \theta_1, \ldots, \theta_n$ in subsequent analysis. Probability and expectation mean conditional probability and expectation, unless otherwise noted.

**3. The bisector test.** An intuitive test for the presence of streaming motion was developed in Mateo, Olszewski and Walker (2008). To begin, stars distant from the center, say, $r \geq r_0$, were selected. There were two reasons for this selection: The effect of streaming motion is expected to be small for stars close to the center and increase with distance from it, and the sample divides into two quite distinct branches for stars at least 400 arc sec from the center; see Figure 2. This resulted in reduced samples which were then divided into two groups by passing bisectors through the data set, and the difference in average velocities for stars in the two groups were computed. The bisectors were of the form $\cos(\theta - \omega) = 0$, where $\omega$ was allowed to vary.

In more detail, let

$$(3) \quad \Delta_0 V(\omega) = \frac{\sum_{r_i>r_0,\cos(\theta_i-\omega)>0} y_i/\sigma_i^2}{\sum_{r_i>r_0,\cos(\theta_i-\omega)>0} 1/\sigma_i^2} - \frac{\sum_{r_i>r_0,\cos(\theta_i-\omega)\leq 0} y_i/\sigma_i^2}{\sum_{r_i>r_0,\cos(\theta_i-\omega)\leq 0} 1/\sigma_i^2},$$

consider the test statistic $B_0 = \max_\omega \Delta_0 V(\omega)$. Attained significance levels, 0.030, 0.006, 0.014, and 0.101 for the reduced samples $r > 400$, $r > 455$, $r > 600$, and $r < 400$, were reported in Mateo, Olszewski, and Walker (2008) using a permutation test.

The idea is sound, but there are details. Supposing that $\nu(r,\theta) = \nu$ is constant, so that there is no streaming motion, let

$$\hat{\nu}_0 = \frac{\sum_{i=1}^n y_i/(\hat{\sigma}_0^2 + \sigma_i^2)}{\sum_{i=1}^n 1/(\hat{\sigma}_0^2 + \sigma_i^2)} \quad \text{and} \quad \hat{\sigma}_0^2 = \frac{1}{n}\sum_{i=1}^n [(y_i - \hat{\nu}_0)^2 - \sigma_i^2].$$

Then $\hat{\nu}_0$ and $\hat{\sigma}_0^2$ are $\sqrt{n}$-consistent estimators of $\nu$ and $\sigma^2$. Define $\Delta_1 V(\omega)$ by (3) with $\sigma_i^2$ replaced by $\hat{\sigma}_0^2 + \sigma_i^2$ and let $B_1 = \max_\omega \Delta_1 V(\omega)$. Using $B_1$ and permuting $(y_1,\sigma_1),\ldots,(y_n,\sigma_n)$ in the permutation test, we obtained somewhat higher significance levels. Plots of $\Delta_1 V(\omega)$ for the reduced samples $r < 400$ and $r > 500$ are shown in Figure 3. For stars at least 500 arc seconds from the center, $\Delta_1 V(\omega)$ attains its maximum value of 10.0682 around 69 degrees, but this spike is very slender, and is, in fact, due to the influence of a single star. One expects the effect of streaming motion to be large along the major axis of Leo I ($\omega = 0$ or 180), and this is the case in Figure 3 and others like it (not included). For this reason, the test that rejects for large values of $|\Delta_1 V(0)|$ is also considered.

Table 2 shows (estimated) significance levels for $B_1$ and $|\Delta_1 V(0)|$ for several values of $r_0$. While the significance levels are higher than reported in Mateo, Olszewski and Walker (2008), they still suggest that streaming motion in present for stars sufficiently far from the center. The dependence on $r_0$ is troubling, however, and the results are far from conclusive. In the next section we present another test which avoids the arbitrary choice of $r_0$, at the expense of setting $\omega = 0$.

TABLE 2
*Test statistics and significance levels*

|  | $B_1$ | $p_{B_1}$ | $\arg\max \Delta_1 V$ | $\|\Delta_1 V(0)\|$ | $p_0$ |
|---|---|---|---|---|---|
| $r < 400$ | 1.5264 | 0.809 | 6.88 | 0.9485 | 0.401 |
| $r > 400$ | 5.9065 | 0.210 | 69.33 | 4.2182 | 0.108 |
| $r > 450$ | 7.5152 | 0.106 | 69.33 | 4.4935 | 0.132 |
| $r > 500$ | 10.0682 | 0.032 | 69.33 | 6.2498 | 0.070 |
| $r > 600$ | 12.2756 | 0.129 | 69.33 | 9.9049 | 0.052 |
| $r > 700$ | 15.0053 | 0.080 | 69.33 | 11.2687 | 0.064 |
| $r > 750$ | 19.4157 | 0.047 | 75.63 | 2.8974 | 0.688 |

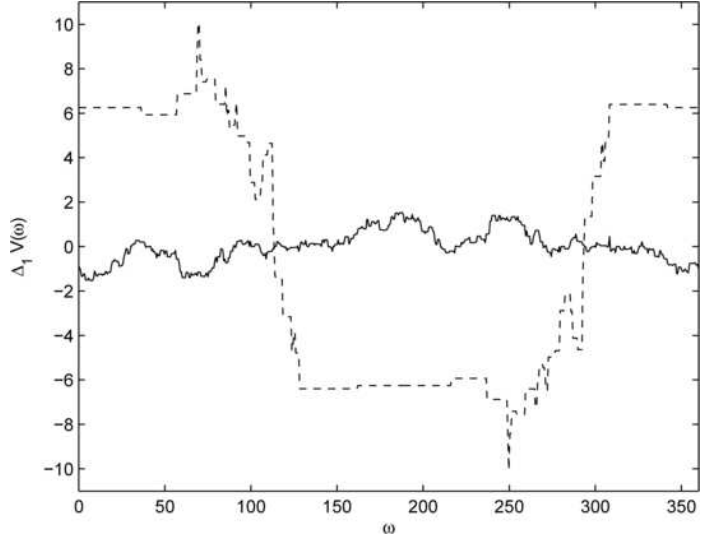

Fig. 3. $\Delta_1 V$ for $r < 400$ (solid) and $r > 500$ (dashed).

The details of the permutation test are as follows. Consider a test statistic $T = T(\mathbf{r}, \boldsymbol{\theta}, \mathbf{y}, \boldsymbol{\sigma})$, where $\mathbf{r} = (r_1, \ldots, r_n)$, $\boldsymbol{\theta} = (\theta_1, \ldots, \theta_n)$, $\mathbf{y} = (y_1, \ldots, y_n)$, and $\boldsymbol{\sigma} = (\sigma_1, \ldots, \sigma_n)$. If there is no streaming motion, then $(R, \Theta)$ and $(Y, \Sigma)$ are independent. In this case $(y_1, \sigma_1), \ldots, (y_n, \sigma_n)$ are conditionally i.i.d. given $(r_1, \theta_1), \ldots, (r_n, \theta_n)$, and the conditional probability that $T > t$ given $(\mathbf{r}, \boldsymbol{\theta})$ and the unordered values $\{(y_1, \sigma_1), \ldots, (y_n, \sigma_n)\}$ is $\#\{\pi : T(\mathbf{r}, \boldsymbol{\theta}, \pi\mathbf{y}, \pi\boldsymbol{\sigma}) > t\}/n!$, where $\pi$ denotes a permutation of $\{1, \ldots, n\}$ and $\pi\mathbf{y}$ and $\pi\boldsymbol{\sigma}$ denote permuted versions of $(y_1, \ldots, y_n)$ and $(\sigma_1, \ldots, \sigma_n)$. Of course, it is not possible to examine all $n!$ permutations, but it is possible to estimate the conditional probability by sampling permutations. The significance levels listed in Table 2, were obtained from 10,000 permutations of the reduced samples. Observe that we permute the pairs $(y_i, \sigma_i)$.

**4. The cosine model.** We now suppose that $\nu(r, \theta) = E(Y | R = r, \Theta = \theta)$ is of the form

$$\nu(r, \theta) = \nu + \lambda(r) \cos(\theta), \tag{4}$$

where $\nu$ is a constant and $\lambda$ is a nonnegative, nondecreasing function. Thus, $|\nu(r, \theta) - \nu|$ is assumed to be nondecreasing in $r$ and largest along the major axis. From an astronomical standpoint, the cosine model is justified by detailed simulations of streaming motions [e.g., Piatek and Pryor (1995),

Muñoz, Majewski and Johnston (2008)]. In these studies the kinematic signature of tidal streaming is noted to closely mimic that of disk rotation. The latter, due to purely geometric projection effects, is precisely modeled by a cosine functional form (with only very minor corrections due to projection effects). Thus, the cosine model is wellmotivated for a wide range of underlying physical mechanisms that might produce the observed kinematics.

*Estimation.* Assuming $\sigma^2$ to be known, the (weighted) conditional least squares estimators $\hat{\nu}$ and $\hat{\lambda}$, given $(\mathbf{r}, \boldsymbol{\theta}, \boldsymbol{\sigma})$, minimize

$$\sum_{i=1}^{n} \frac{[y_i - v - u(r_i)\cos(\theta_i)]^2}{\sigma^2 + \sigma_i^2},$$

with respect to $v \in \mathbb{R}$ and nonnegative, nondecreasing functions $u$. Differentiation then gives the following conditions for the least squares estimators:

$$\sum_{i=1}^{n} \frac{y_i - \hat{\nu} - \hat{\lambda}(r_i)\cos(\theta_i)}{\sigma^2 + \sigma_i^2} = 0$$

and

$$\sum_{i=1}^{n} \cos(\theta_i) \left( \frac{y_i - \hat{\nu} - \hat{\lambda}(r_i)\cos(\theta_i)}{\sigma^2 + \sigma_i^2} \right) \xi_i \leq 0$$

for all nonnegative, nondecreasing $0 \leq \xi_1 \leq \cdots \leq \xi_n$. We will use these conditions with $\sigma^2$ replaced by

(5) $$\hat{\sigma}^2 = \frac{1}{n} \sum_{i=1}^{n} \{[y_i - \hat{\nu} - \hat{\lambda}(r_i)\cos(\theta_i)]^2 - \sigma_i^2\}.$$

Thus, letting $\hat{w}_i = 1/(\hat{\sigma}^2 + \sigma_i^2)$ and $\hat{W}_n = \hat{w}_1 + \cdots + \hat{w}_n$,

(6) $$\hat{\nu} = \frac{1}{\hat{W}_n} \sum_{i=1}^{n} \hat{w}_i [y_i - \hat{\lambda}(r_i)\cos(\theta_i)]$$

and

(7) $$\hat{\lambda}(r_k) = \max[0, \tilde{\lambda}(r_k)],$$

where

(8) $$\tilde{\lambda}(r_k) = \max_{i \leq k} \min_{j \geq k} \frac{\hat{w}_i \cos(\theta_i)(y_i - \hat{\nu}) + \cdots + \hat{w}_j \cos(\theta_j)(y_j - \hat{\nu})}{\hat{w}_i \cos^2(\theta_i) + \cdots + \hat{w}_j \cos^2(\theta_j)}.$$

Alternatively, letting $\hat{T}_k = \hat{w}_1 \cos^2(\theta_1) + \cdots + \hat{w}_k \cos^2(\theta_k)$,

$$\Lambda^\#(t) = \sum_{i: \hat{T}_i \leq t} \hat{w}_i \cos(\theta_i)[y_i - \hat{\nu}],$$

and $\tilde{\Lambda} = \mathrm{GCM}(\Lambda^{\#})$, the greatest convex minorant of $\Lambda^{\#}$, $\tilde{\lambda}(r_k) = \tilde{\Lambda}'(\hat{T}_k)$, the left-hand derivative of $\tilde{\Lambda}$ at $\hat{T}_k$. See Robertson, Wright and Dykstra (1988), Chapter 1 for background on isotonic estimation.

For Leo I, iterating (5), (6), and (7) leads to convergence to three decimal places after four iterations. For this data set, $\hat{\sigma} = 9.0107$ and $\hat{\nu} = 283.1040$. The function $\hat{\lambda}$ is graphed in the left panel of Figure 4. The large value at the right end point is almost certainly due to the spiking problem: Isotonic estimators of an increasing function are inconsistent at the right end point, where they are simply too large. Heuristically, the reason for this can be seen from (8): $\tilde{\lambda}(r_n)$ is the maximum of an increasing number of terms (if $\hat{\sigma}$ and $\hat{\nu}$ are held fixed). A similar, but more severe, problem arises in monotone density estimation and has received more attention. See Woodroofe and Sun (1993) and Kulikov and Lopuhaä (2006). To eliminate spiking, we replace $\hat{\lambda}(r_n)$ by 6.193, the average of the last thirteen values of $\hat{\lambda}(r_k)$, adapting the suggestion of Kulikov and Lopuhaä (2006) to our context where data are much sparser. This limits the effect of the last observation in the subsequent calculations. The truncated $\hat{\lambda}$ appears in the right panel of Figure 4.

*Confidence intervals.* The asymptotic distribution of $\hat{\lambda}(r)$ can be derived under modest conditions. Suppose that $\lambda$ is continuously differentiable around $r > 0$, $\lambda'(r) > 0$, and let $C = E[1/(\sigma^2 + \Sigma^2)]$ and

$$\gamma_r = 2 \left| \frac{\lambda'(r)}{2C \int f(r,\theta) \cos^2(\theta) \, d\theta} \right|^{1/3}.$$

Then, a (fairly) straightforward application of the Argmax theorem [van der Vaart and Wellner (2000)] shows that the asymptotic unconditional distribution of $\mathbb{C}_n = n^{1/3}[\hat{\lambda}(r) - \lambda(r)]/\gamma_r$ is Chernoff's distribution [Groeneboom and Wellner (2001)], the distribution of $\arg\min_t \mathbb{W}(t) + t^2$, where $\mathbb{W}$ denotes a standard two sided Brownian motion. Thus, $\mathbb{C}_n$ is an asymptotic pivot, but it is difficult to use this result to set confidence intervals for $\lambda(r)$, because it is difficult to estimate $\lambda'(r)$ and hence the normalizing constant $\gamma_r$. Moreover, even the condition $\lambda'(r) > 0$ is suspect on the interval where $\hat{\lambda} = 0$.

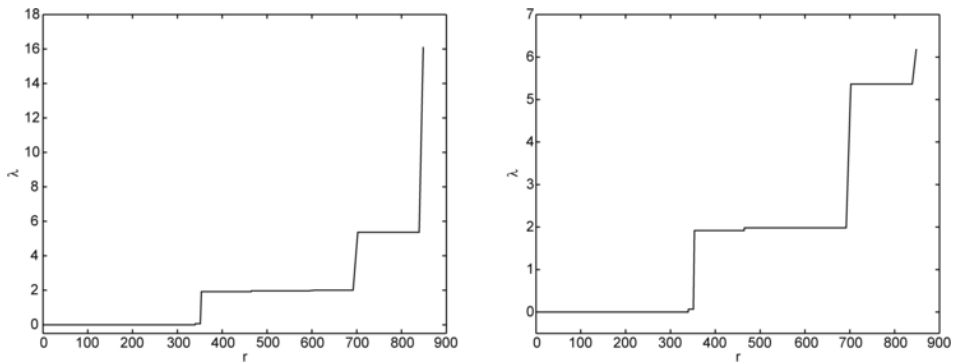

FIG. 4. *Left:* $\hat{\lambda}$ *before truncation. Right:* $\hat{\lambda}$ *after truncation.*

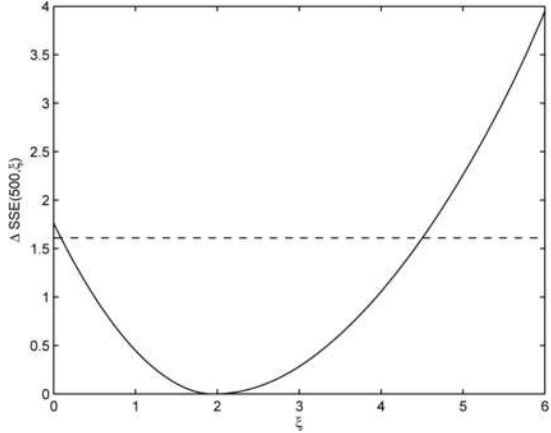

FIG. 5. $\Delta SSE(500, \xi)$ as $\xi$ varies from 0 to 6, with the 90% cut-off mark (dashed).

It is possible to avoid the problem of estimating $\gamma_r$, though not the condition $\lambda'(r) > 0$, by adapting the likelihood based confidence intervals of Banerjee and Wellner (2001) to the present problem. For fixed $r_0, \xi_0 > 0$, and $\sigma > 0$, let

$$\Delta SSE(r_0, \xi_0) = \min_{u(r_0) = \xi_0} \sum_{i=1}^{n} \frac{[y_i - v - u(r_i)\cos(\theta_i)]^2}{\sigma^2 + \sigma_i^2} \tag{9}$$
$$- \min \sum_{i=1}^{n} \frac{[y_i - v - u(r_i)\cos(\theta_i)]^2}{\sigma^2 + \sigma_i^2},$$

where implicitly both minimizations are over $v \in \mathbb{R}$ and nonnegative, nondecreasing functions $u$. If $\lambda(r_0) = \xi_0$, $\lambda$ is continuously differentiable near $r_0$, and $\lambda'(r_0) > 0$, then $\Delta SSE(r_0, \xi_0)$ has a limiting distribution that does not depend on any unknown parameters; and the same asymptotic distribution is obtained with $\sigma$ replaced by $\hat{\sigma}$, as $\hat{\sigma}$ converges at a much faster rate. A description of the asymptotic distribution, including graphs and percentiles, may be found in Banerjee and Wellner (2005). In particular, (Monte Carlo estimates of) the 90th and 95th percentile are 1.61 and 2.29 and, for example, $\{\xi : \Delta SSE(r_0, \xi) \leq 1.61\}$ is an approximate 90% confidence set for $\lambda(r_0)$. A plot of $\Delta SSE(500, \xi)$ is shown in Figure 5, and selected confidence intervals are listed in Table 3.

Resampling provides still another way to set confidence intervals. Recent results [Kosorok (2007), Lee and Pun (2006), Léger and MacGibbon (2006),

Sen, Banerjee and Woodroofe (2008)], some obtained for related problems, suggest that direct use of the bootstrap will not provide consistent estimators of the distribution of sampling error but that use of a smoothed bootstrap or $m$ out of $n$ bootstrap will. Thus, let

$$\hat{\lambda}_s(e^t) = \int_{-\infty}^{\infty} \hat{\lambda}(e^u) \frac{1}{b} K\left(\frac{u-t}{b}\right) du,$$

where $K$ is a kernel and $b$ a bandwidth. We used the standard normal density for $K$ and chose the bandwidth $b = 0.1$ (subjectively) to compromise between smoothness and fit. The result is shown in the left panel of Figure 6. The right panel shows the derivative of the smoothed estimator, illustrating the difficulty in estimating $\lambda'(r)$.

Now, let $e_i$ denote the residuals, $e_i = y_i - \hat{\nu} - \hat{\lambda}_s(r_i)\cos(\theta_i)$, $i = 1,\ldots,n$. Let $x_i = (\sigma_i, e_i)$, and let $F^\#$ denote the empirical distribution of $x_1,\ldots,x_n$. Further, let $(S_1, Z_1),\ldots,(S_n, Z_n) \sim F^\#$ be conditionally independent given $(\mathbf{r}, \boldsymbol{\theta}, \mathbf{y}, \boldsymbol{\sigma})$; let

(10) $$y_i^* = \hat{\nu} + \hat{\lambda}_s(r_i)\cos(\theta_i) + Z_i \quad \text{and} \quad \sigma_i^* = S_i;$$

and let $\hat{\lambda}^*$ denote the (truncated) isotonic estimator (7) computed from $y_1^*,\ldots,y_n^*$ and $\sigma_1^*,\ldots,\sigma_n^*$ with $r_1,\ldots,r_n$ and $\theta_1,\ldots,\theta_n$ held fixed. To set confidence intervals for $\lambda(r_0)$, we estimate the distribution of $\hat{\lambda}(r_0) - \lambda(r_0)$ by the conditional distribution of $\hat{\lambda}^*(r_0) - \hat{\lambda}_s(r_0)$, which may be computed

TABLE 3
90% and 95% confidence intervals for $\lambda(r)$

| | $\Delta SSE$ | | | | | | Bootstrap | | | | |
|---|---|---|---|---|---|---|---|---|---|---|---|
| | 0.90 | | | 0.95 | | | 0.90 | | 0.95 | | |
| $r_0$ | L | U | CP | L | U | CP | L | U | L | U | $\hat{\lambda}$ |
| 400 | 0 | 3.54 | 0.901 | 0 | 3.86 | 0.952 | 0 | 3.57 | 0 | 3.57 | 1.92 |
| 450 | 0 | 3.63 | 0.887 | 0 | 4.01 | 0.943 | 0 | 3.74 | 0 | 3.85 | 1.92 |
| 500 | 0.10 | 4.50 | 0.882 | 0 | 5.02 | 0.936 | 0 | 3.58 | 0 | 3.90 | 1.98 |
| 550 | 0.26 | 4.65 | 0.852 | 0 | 5.19 | 0.916 | 0 | 3.44 | 0 | 3.76 | 1.98 |
| 600 | 0.26 | 6.66 | 0.827 | 0 | 7.30 | 0.897 | 0 | 3.30 | 0 | 3.64 | 1.98 |
| 650 | 0.36 | 6.70 | 0.865 | 0.05 | 7.39 | 0.922 | 0 | 3.58 | 0 | 3.97 | 1.99 |
| 700 | 0.36 | 8.88 | 0.913 | 0.05 | 9.56 | 0.961 | 0 | 4.26 | 0 | 4.69 | 1.99 |
| 750 | 1.85 | 8.88 | 0.906 | 1.37 | 9.56 | 0.952 | 0.44 | 7.86 | 0 | 8.37 | 5.37 |

*Notes*: The leftmost column shows the radial distance. The next two columns are lower and upper endpoints of an approximate 90% confidence interval computed from $\Delta SSE$; the fourth column is a bootstrap estimate of the coverage probability; the fifth, sixth, and seventh columns provide the same information for 95% confidence intervals. The next four columns are lower and upper endpoints of approximate 90% and 95% confidence intervals computed from the bootstrap. The last column provides the value of $\hat{\lambda}$.

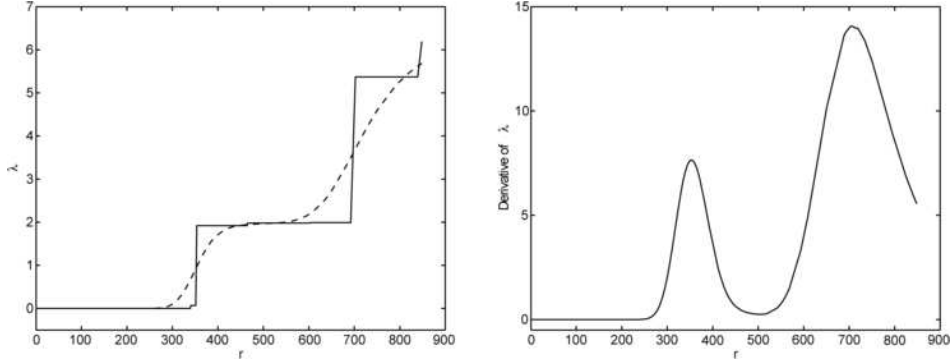

FIG. 6. *Left: The smoothed (dashed) and unsmoothed (solid) estimators with $b = 0.1$. Right: The derivative of the smoothed estimator.*

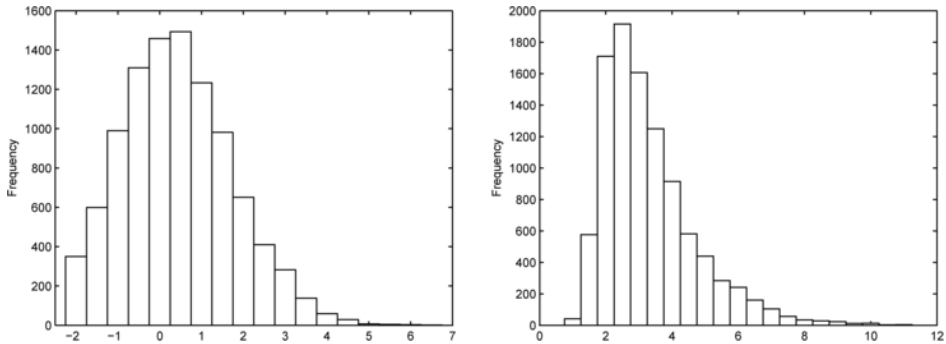

FIG. 7. *Histograms of $10,000$ realizations of $\hat{\lambda}^*(500) - \hat{\lambda}_s(500)$ (left) and $D^*$ (right).*

from simulation. The left panel in Figure 7 shows a histogram of 10,000 values of $\hat{\lambda}^*(500) - \hat{\lambda}_s(500)$. Bootstrap confidence intervals for selected $r_0$ are listed in Table 3. Similarly, to set confidence bands for $\lambda$, we approximate the distribution of $D = \max_r |\hat{\lambda}(r) - \lambda(r)|$ by that of $D^* = \max_r |\hat{\lambda}^*(r) - \hat{\lambda}_s(r)|$. The right panel in Figure 7 shows a histogram of 10,000 values of $D^*$. The 90th and 95th percentiles of this distribution are 5.194 and 6.074. We have not standardized these variables before bootstrapping, because it is difficult to estimate $\lambda'$.

Unfortunately, there are major differences between the two methods for setting confidence intervals. To some extent, these can be explained by the constructions: The bootstrap intervals attempt to balance the error probabilities equally, left and right; the intervals derived from $\Delta SSE$ make no such attempt. There are more serious differences, however, between the asymptotic values and the bootstrap estimates. The difference can be seen in the left panel of Figure 7: The histogram is asymmetric, whereas Chernoff's distribution is symmetric. The $\Delta SSE$ method depends on the ap-proximations $P\{\Delta SSE[r_0, \lambda(r_0)] \leq 1.61\} \approx 0.90$ and $P\{\Delta SSE[r_0, \lambda(r_0)] \leq 2.29\} \approx 0.95$, where now $P$ denotes the unconditional probability. The bootstrap estimates of these probabilities, $P^*\{\Delta SSE^*[r_0, \hat{\lambda}_s(r_0)] \leq 1.61\}$ and $P^*\{\Delta SSE^*[r_0, \hat{\lambda}_s(r_0)] \leq 2.29\}$, are reported in columns three and six of Table 3. There is good agreement for $r_0 \leq 450$ and $r_0 \geq 700$. This is important, because the positive lower confidence bounds on the last two lines of Table 3 reinforce the conclusions in Section 3 that there is streaming motion. But the bootstrap estimates are substantially less than the nominal values for $550 \leq r \leq 650$, an interval that includes values for which $\hat{\lambda}'_s(r_0)$ is very small, and the positive lower confidence limits in this region should not be trusted. The disagreement between the bootstrap intervals and ones derived from asymptotic distributions is difficult to resolve, in part, because the justification for the bootstrap is itself asymptotic and requires the condition $\lambda'(r_0) > 0$.

*Testing.* Within the cosine model (4), testing for streaming motion means testing the null hypothesis $\lambda = 0$. The positive lower confidence limits on the last lines of Table 3 suggest that this hypothesis can be rejected. This point can be made in another way that does not depend on asymptotics or even the validity of (4). Consider the F-like statistic

$$\mathcal{F} = \sum_{i=1}^{n} \hat{w}_i \cos^2(\theta_i) \hat{\lambda}^2(r_i), \tag{11}$$

were known, $\varepsilon_1, \ldots, \varepsilon_n$ were normal, $\hat{w}_i$ were replaced by $w_i = 1/(\sigma^2 + \sigma_i^2)$, and $\hat{\lambda}$ were replaced by the isotonic estimator for known $\nu$ and $\sigma^2$, then $-2\mathcal{F}$ would be the log-likelihood ratio statistic for testing $\lambda = 0$; see, Chapter 2 of Robertson, Wright and Dykstra (1988).

For the Leo I data set, the observed value of $\mathcal{F}$ was 6.69. We again assess significance from the permutation distribution of $\mathcal{F}$, but computed from the full sample $(r_i, \theta_i, y_i, \sigma_i), i = 1, \ldots, n = 328$. In a sample of 10,000 permutations, the permuted value of $\mathcal{F}$ exceeded the observed value 553 times, roughly confirming the conclusion based on confidence intervals and asymptotic calculations. The effect of truncation on the test statistic is amusing. Without the truncation the value of $\mathcal{F}$ would have been substantially larger, 8.72, but the significance level would have been essential unchanged, 0.0543.

Observe that (4) was used only to motivate the form of the test statistic. The F-like test statistic serves also as a test of the hypothesis $\nu(r, \theta) = \nu$, in (2).

**5. Thresholds and the break point.** By a threshold or breakpoint we mean a distance from the center of Leo I below which there is no streaming motion, or very little, and above which streaming motion is appreciable. This threshold or tidal radius of the dwarf galaxy corresponds to the minimum radius where stars remain bound. Outside this radius, stars previously bound to the dwarf galaxy are formally on independent orbits about the Milky Way, though of course these orbits are generally very similar to that of the dwarf galaxy itself. This produces the streaming motion our method aims to detect. Consequently, one can expect that the kinematic streaming effects will only begin to appear outside the tidal radius, a feature our threshold model aims to mimic. Of course, projection effects and the fact that the stars of any galaxy exhibit a distribution of velocities mean that the threshold effect will not be arbitrarily sharp, but the transition from nonstreaming well inside the tidal radius to streaming outside this radius is quite reasonably motivated in an astronomical context. We consider two approaches to defining and estimating such a threshold, change point models and split points, as in Banerjee and McKeague (2007a).

*Change point models.* Let $\tau$ denote an upper limit for $r$. In the change point model it is assumed that there is a $\rho > 0$ for which $\lambda(r) = 0$ for $r \leq \rho$ and $\lambda(r) > 0$ for $\rho < r \leq \tau$, in which case we call $\rho$ the *threshold*.

We may obtain an upper confidence bound by modifying the F-like statistic (11). For a given $\rho_0$ consider the hypothesis $\rho \geq \rho_0$. Let

$$\mathcal{F}(\rho_0) = \sum_{i: r_i \leq \rho_0} \hat{w}_i \cos^2(\theta_i) \hat{\lambda}^2(r_i)$$

and let $m$ be the largest integer for which $r_m \leq \rho_0$. Then the conditional null distribution of $\mathcal{F}(\rho_0)$ is invariant under permutations of $(y_1, \sigma_1), \ldots, (y_m, \sigma_m)$, so that significance can again be assessed from a permutation distribution of $\mathcal{F}(\rho_0)$. Of course, the set of $\rho_0$ for which the hypothesis is accepted at level $\alpha$ is a level $1 - \alpha$ confidence set for $\rho$. For the Leo I data set with $1 - \alpha = 0.9$, this hypothesis is rejected when $\rho_0 = 720$, which, therefore, serves as an upper confidence bound. Again, the cosine model (4) was used only to motivate the form of the test statistic. The test just described also serves as a test of $\nu(r, \theta) \equiv \nu$ for all $\theta$ and all $r \leq \rho_0$.

Unsurprisingly, adopting an even more structured model suggests a lower bound. Suppose that $\lambda(r) = \beta \psi(r - \rho)$, where $\beta > 0$ is an unknown parameter and $\psi$ is a known function for which $\psi(x) = 0$ for $x < 0$ and $\psi(x) > 0$ for $x > 0$. Thus,

$$y_i = \nu + \beta \psi(r_i - \rho) \cos(\theta_i) + \varepsilon_i + \delta_i \tag{12}$$

for $i = 1, \ldots, n$. For a fixed $\rho$ this is a simple linear regression model. Let $\hat{\beta}_r$ and $\hat{\nu}_r$ denote the weighted least squares estimators, using the weights $\hat{w}_i = 1/(\hat{\sigma}^2 + \sigma_i^2)$, derived from (12) assuming $\rho = r$, and let $SSE_r$ denote the residual sum of squares,

$$SSE_r = \sum_{i=1}^n \hat{w}_i [y_i - \hat{\nu}_r - \hat{\beta}_r \psi(r_i - r) \cos(\theta_i)]^2.$$

Then the LSE $\hat\rho$ of $\rho$ minimizes $SSE_r$ with respect to $r$. Figure 8 shows graphs of $SSE_r - SSE_{\hat\rho}$ for two choices of $\psi$, $\psi(x) = \mathbf{1}_{(0,\infty)}(x)$ and $\psi(x) = \max(0,x)$. The latter choice leads to the segmented regression model considered in Hinkley (1971), Feder (1975), and recently in Hušková (1998). For this choice it may be shown that $SSE_\rho - SSE_{\hat\rho}$ has a limiting $\chi_1^2$ distribution, assuming (12). So, $\{r : SSE_r - SSE_{\hat\rho} \leq c\}$ is an asymptotic level $P[\chi_1^2 \leq c]$ confidence set for $\rho$. For Leo I, the 90% asymptotic confidence set $[0, 654.5] \cup [823.1, 848.5]$ so obtained is disconnected, but there are only four stars for which $823.1 \leq r \leq 848.5$, and this interval is of little interest. Letting $\psi(x) = \mathbf{1}_{(0,\infty)}(x)$ leads to (a minor variation on) the classical change point problem. The asymptotic distribution of $SSE_\rho - SSE_{\hat\rho}$ may be obtained for this case too, but it is complicated and unnecessary in the sense that $SSE_r - SSE_{\hat\rho}$ rises and falls so sharply near $r = 333.5$ and $r = 702.7$.

*Split points.* It is possible to define and estimate a breakpoint without assuming that $\lambda(r)$ is actually equal to 0 for small $r$, by fitting a stump model $\beta \mathbf{1}_{(\gamma,\tau]}$ to $\lambda$, as in Banerjee and McKeague (2007a). This is accomplished by minimizing

$$\kappa(b,r) = \int_0^r \lambda^2(s) h(s)\, ds + \int_r^\tau [\lambda(s) - b]^2 h(s)\, ds,$$

with respect to $b$ and $r$. Here $h$ is a positive continuous weight function. We used $h = 1$ in Figure 9. Another possibility is to let $h$ be the marginal density of $R$, which is known from Irwin and Hatzidimitriou (1995). The minimization with respect to $b$, of course, is simple, and

$$(13) \qquad \kappa_0(r) := \min_b \kappa(b,r) = \int_0^\tau \lambda^2(s) h(s)\, ds - \frac{[\Lambda(\tau) - \Lambda(r)]^2}{H(\tau) - H(r)},$$

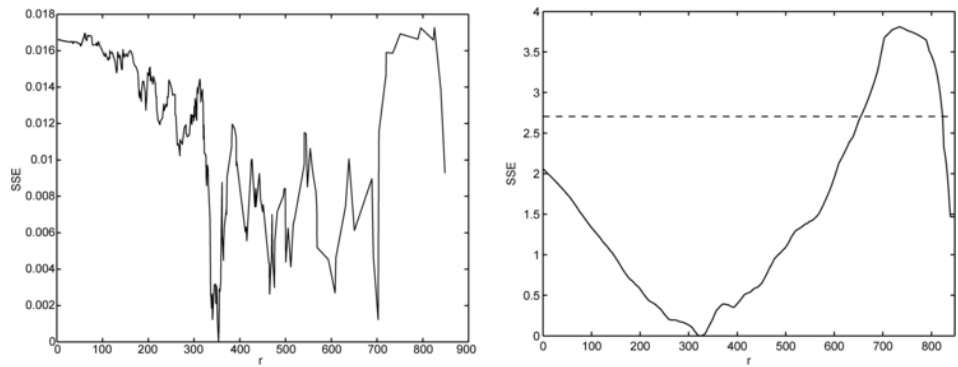

Fig. 8. *Graphs of $SSE_r - SSE_{\hat\rho}$ for $\psi(x) = \mathbf{1}_{(0,\infty)}(x)$ (left) and $\psi(x) = \max(0,x)$ (right).*

where $H(r) = \int_0^r h(s)\,ds$ and $\Lambda(r) = \int_0^r \lambda(s)h(s)\,ds$. We define the *break point* $\gamma$ to be the minimizing value of $r$, or the infimum of minimizing values if there is more than one. If $\lambda$ is continuous and $\lambda(0) < \Lambda(\tau)/[2H(\tau)]$, then

$$\lambda(\gamma) = \frac{[\Lambda(\tau) - \Lambda(\gamma)]}{2[H(\tau) - H(\gamma)]}.$$

Observe that if there is a threshold $\rho$ [for which $\lambda(r) = 0$ for $r \leq \rho$], then $\gamma \geq \rho$, because $\Lambda(r) = 0$ for $r \leq \gamma$.

The simplest way to estimate $\gamma$ is to replace $\lambda$ by $\hat{\lambda}$ in the definition. The left panel of Figure 9 shows a graph of $\hat{\kappa}_{00}(r)$, where $\hat{\kappa}_{00}$ denotes $\kappa_0$ with $\lambda$ replaced by $\hat{\lambda}$ and rescaled to take values between 0 and 1. That is, letting $\hat{\kappa}_0$ denote $\kappa_0$ with $\lambda$ replaced by $\hat{\lambda}$ and $\hat{\gamma}$ a minimizing value of $\hat{\kappa}_0(r)$,

$$\hat{\kappa}_{00}(r) = \frac{\hat{\kappa}_0(r) - \hat{\kappa}_0(\hat{\gamma})}{\max_s \hat{\kappa}_0(s) - \hat{\kappa}_0(\hat{\gamma})}.$$

With $h \equiv 1$, the estimated break point is 353.5 arc sec, but there is a near minimum at about 700.

Asymptotic theory is not uninteresting from a theoretical point of view, but provides little useful guidance here. As described in the Appendix, the asymptotic unconditional distribution of $\hat{\gamma}$ depends on $\lambda'(\gamma)$ and would have to be approximated by simulation in any case. A bootstrap procedure does provide some guidance, however. Define $y_i^*$ and $\sigma_i^*$ as in (10); let $\hat{\gamma}_s$ denote the value of $\gamma$ obtained by replacing $\lambda$ by $\hat{\lambda}_s$; and let $\hat{\gamma}^*$ and $\hat{\kappa}_{00}^*$ denote the values of $\hat{\gamma}$ and $\hat{\kappa}_{00}$ respectively, computed from $y_1^*, \ldots, y_n^*, \sigma_1^*, \ldots, \sigma_n^*$, with $r_1, \ldots, r_n, \theta_1, \ldots, \theta_n$ held fixed. Then the conditional distribution of $\hat{\kappa}_{00}^*(\hat{\gamma}_s)$ provides an estimate of the distribution of $\hat{\kappa}_{00}(\gamma)$. A histogram of 10,000 values of $\hat{\kappa}_{00}^*(\hat{\gamma}_s)$ is shown in the right panel of Figure 9. The 90th and 95th percentiles of this distribution are 0.3694 and 0.4690. So, for example, the set of $r$ for which $\hat{\kappa}_{00}(r) \leq 0.3694$ is a 90% bootstrap confidence set for $\gamma$. Unfortunately, this is a large interval, $[91.1, 734.3]$.

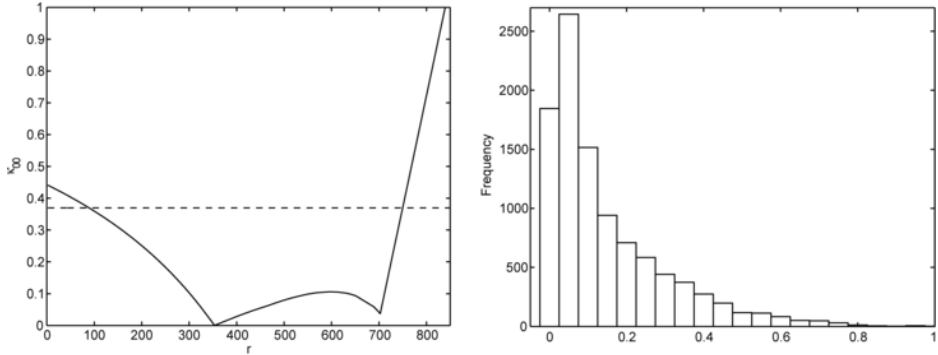

FIG. 9. *Left:* $\hat{\kappa}_{00}(r)$. *Right: Bootstrap estimate of the distribution of* $\hat{\kappa}_{00}(\gamma)$.

## 6. Some remarks.

1. The conclusions regarding the presence of streaming motion have to be tentative, because of the large significance levels. One of the key factors behind this is the comparatively small sample size ($n = 328$) and, in particular, the very few observations far out from the center of the galaxy, the region of interest. In fact, we just have 64 data points above 400 arc seconds. We hope that with more data in the future our methods can be used more effectively to draw stronger conclusions. We also expect to obtain data on other dwarf spheroidal galaxies, for example, Draco, Fornax, etc., and will be applying variants of our methods to analyze the samples.
2. With more data we can resort to more flexible modeling. For example, instead of simply using $\cos\theta$, we could model the effect of angle $\theta$, by a function $h(\cos\theta)$ or $h(\cos(\theta - \omega))$, where $h$ is nondecreasing and $\omega$ is an unknown parameter representing the direction of tidal streaming. The velocity dispersion parameter, $\sigma^2$, has been assumed to be constant in our approach. For Leo I, there is evidence for such an assumption [see Mateo, Olszewski and Walker (2008)]. For other galaxies, it is conceivable that $\sigma$ may depend on $r$, the radial distance from the center of the galaxy. Our methods should be adaptable to this heteroscedasticity, but may require nontrivial extensions.
3. Monotone regression splines provide a method for combining monotonicity constraint and smoothness. These were used effectively in Wang et al. (2005, 2008a, 2008b) and could be investigated in the present context.
4. The smoothed bootstrap was invoked at several places in this paper for purposes of uncertainty assessment—for example, in constructing pointwise and simultaneous confidence bands for $\lambda$, and confidence sets for the split point $\gamma$. While there is evidence [see Kosorok (2007), Sen, Banerjee and Wellner (2008) and Léger and McGibbon (2006)] that the smoothed bootstrap provides consistent estimates of pointwise confidence sets for $\lambda$, the use of this method for approximating the distribution of $D$ and that of $\hat{\kappa}_{00}(\gamma)$ remains to be vindicated. In particular, little is known about the limiting behavior of $D$ or $\hat{\kappa}_{00}(\gamma)$. An alternative to the smoothed bootstrap would have been to use subsampling or the $m$ out of $n$ bootstrap.
5. The asymptotic distribution of the least squares estimate of $\gamma$ in the split point model, stated in the Appendix, is curious in the sense that (a multiple of) Chernoff's distribution no longer arises and is replaced by a nonstandard limit. We know of no other situations in the published literature where this limit distribution has been encountered. Furthermore, it does not seem possible to represent the limit as a multiple of a universal distribution, which renders the computation of quantiles difficult.

6. An interesting but difficult problem is to estimate the threshold parameter $\rho$ nonparametrically, assuming only that $\lambda(r) = 0$ for $r \leq \rho$, $\lambda(r) > 0$ for $r > \rho$ and that $\lambda$ is increasing. We expect the rate at which $\rho$ can be estimated to depend crucially on the smoothness of the join between the two segments of the function at $\rho$; the smoother the join, the slower the convergence. The intuitive estimator $\inf\{t : \hat{\lambda}(t) > 0\}$ underestimates $\rho$ heavily. One suggested modification is to replace 0 by a positive threshold that decreases to 0 at an appropriate rate. Yet another approach would be to construct a penalized least squares estimate of $\lambda$ under monotonicity constraints, where one penalizes monotone functions with low values of the threshold parameter.

7. Yet another way of estimating $\gamma$ is to observe that $(\nu, \beta, \gamma) = \arg\min_{(v,b,r)} \mathbb{M}(v,b,r)$ and

$$\mathbb{M}(v,b,r) = E\left[\{Y - v - b\mathbf{1}(R > r)\cos\Theta\}^2 \frac{1}{\phi(R, \Theta, \Sigma)}\right],$$

with $\phi(R,\Theta,\Sigma) = f(R,\Theta)(\sigma^2 + \Sigma^2)/h(R)$. We can approximate this criterion function $\mathbb{M}$, by the empirical expectation, and construct an estimate of $\gamma$ as the threshold that minimizes the sample analogue. This method avoids the estimation of $\lambda$. However, it needs knowledge of $\phi$, which, in turn, involves estimation of $f$. This is not feasible with the currently available sample size, so we have not explored this approach in the paper.

## APPENDIX

The Appendix contains descriptions of two asymptotic distributions that arise in the paper. The first of these may be derived by a straightforward adaptation of the arguments in Banerjee (2007) and Banerjee and Wellner (2001). The derivation of the second will appear elsewhere.

**The residual sum of squares statistic $\Delta SSE(r_0, \xi_0)$.** For a real-valued function $f$ that is bounded below on $\mathbb{R}$, let $\text{gcm}(f, I)$ and $\text{slogcm}(f, I)$ denote the greatest convex minorant (GMC) and the left–hand slope of the GCM of the restriction of $f$ to the interval $I$. We abbreviate $\text{gcm}(f, \mathbb{R})$ and $\text{slogcm}(f, \mathbb{R})$ by $\text{gcm}(f)$ and $\text{slogcm}(f)$ and let

$$\text{slogcm}^0(f) = (\text{slogcm}(f, (-\infty, 0]) \wedge 0)\mathbf{1}_{(-\infty, 0]}$$
$$+ (\text{slogcm}(f, (0, \infty)) \vee 0)\mathbf{1}_{(0, \infty)}.$$

For positive constants $c$ and $d$ define the process $\mathbb{X}_{c,d}(t) = c\mathbb{W}(t) + dt^2$, where $\mathbb{W}$ is, standard two-sided Brownian motion starting from 0. Let $g_{c,d} = \text{slogcm}(\mathbb{X}_{c,d})$ and $g^0_{c,d} = \text{slogcm}^0(\mathbb{X}_{c,d})$. Thus, $g_{1,1}$ and $g^0_{1,1}$ are the unconstrained and constrained versions of the slope processes associated with the "canonical" process $\mathbb{X}_{1,1}$. For details about the processes $g_{c,d}$ and $g_{c,d}^0$, see Banerjee (2007) and Banerjee and Wellner (2001). Set

$$\mathbb{D}_{c,d} = \int [\{g_{c,d}(u)\}^2 - \{g_{c,d}^0(u)\}^2] \, du$$

and abbreviate $\mathbb{D}_{1,1}$ to $\mathbb{D}$. Then $\mathbb{D}_{c,d}$ has the same distribution as $c^2 \mathbb{D}$ [see Banerjee (2007) and Banerjee and Wellner (2001)]. With this notation, *if $\lambda(r_0) = \xi_0$, $\lambda$ is continuously differentiable near $r_0$, and $\lambda'(r_0) > 0$, then $\Delta SSE(r_0, \xi_0)$ converges in distribution to $\mathbb{D}$.* This may be established by adapting the arguments of Banerjee and Wellner (2001) to the cosine model.

**The split point estimation procedure.** Recall the definitions of the *break-point* $\gamma$ and the best fitting $\beta$ from Section 5 along with those of $H(r) = \int_0^r h(s) \, ds$ and $\Lambda(r) = \int_0^r \lambda(s)h(s) \, ds$. Thus, $\beta = [\Lambda(\tau) - \Lambda(\gamma)]/[H(\tau) - H(\gamma)]$. Next let $\mathbb{G}_{c,d} = \text{gcm}(\mathbb{X}_{c,d})$,

$$a = \sqrt{\frac{1}{E[(\sigma^2 + \Sigma^2)^{-1}] \int \cos^2(\theta) f(r_0, \theta) \, d\theta}}, \qquad b = \frac{1}{2}\lambda'(r_0),$$

$$V = \begin{bmatrix} 2[H(\tau) - H(\gamma)] & -2\lambda(\gamma)h(\gamma) \\ -2\lambda(\gamma)h(\gamma) & 4\lambda(\gamma)\lambda'(\gamma)h(\gamma) \end{bmatrix}$$

and

$$\mathbb{Q}(t_1, t_2) = 2\beta h(\gamma)[\mathbb{G}_{a,b}(t_2) - \mathbb{G}_{a,b}(0) - bt^2] + \tfrac{1}{2}\mathbf{t}'V\mathbf{t},$$

with $\mathbf{t} = (t_1, t_2)$. Now let $\hat{\gamma}$ and $\hat{\beta}$ denote the estimators of $\gamma$ and $\beta$ obtained by replacing $\lambda$ by $\hat{\lambda}$. *If* (i) *$\gamma$ is the unique minimizer of $\kappa_0$ in (13),* (ii) *$\lambda$ is continuously differentiable near $\gamma$, $\lambda'(\gamma) > 0$, and* (iii) *$(\hat{\beta}, \hat{\gamma})$ converges to $(\beta, \gamma)$ at the rate $n^{-1/3}$, then $n^{1/3}(\hat{\beta} - \beta, \hat{\gamma} - \gamma)$ converges in distribution to* $\text{argmin}_\mathbf{t} \mathbb{Q}(\mathbf{t})$. Here assumptions (i) and (ii) seem quite reasonable and even necessary in the sense that $\lambda'(\gamma)$ appears in the asymptotic distribution. Assumption (iii) is inelegant. It is suggested by Banerjee and McKeague (2007b), but a complete proof is still lacking in the present context.

The asymptotic distribution of $n^{1/3}(\hat{\gamma} - \gamma)$ may be simplified by first minimizing $\mathbb{Q}(t_1, t_2)$ with respect to $t_1$. After some algebra, it is found to be the distribution of $\text{argmin}_{t_2}[\mathbb{P}_{a,b}(t_2) + ct_2^2]$, where $\mathbb{P}_{a,b}(t_2) = \mathbb{G}_{a,b}(t_2) - bt_2^2$ and

$$c = \frac{1}{2}\bigg[\lambda'(\gamma) - \frac{1}{2}\bigg(\frac{\lambda(\gamma)h(\gamma)}{H(\tau) - H(\gamma)}\bigg)\bigg].$$

B. Sen  
M. Banerjee  
M. Woodroofe  
Department of Statistics  
University of Michigan  
1085 South University  
Ann Arbor, Michigan 48109  
USA  
E-mail: bodhi@umich.edu  
    moulib@umich.edu  
    michaelw@umich.edu  

M. Mateo  
Department of Astronomy  
University of Michigan  
Ann Arbor, Michigan 48109-1042  
USA  
E-mail: mmateo@umich.edu  

M. Walker  
Institute of Astronomy  
University of Cambridge  
Madingley Road  
Cambridge CB3 0HA  
United Kingdom  
E-mail: walker@ast.cam.ac.uk